\documentclass[twocolumn]{article}

\usepackage{arxiv}

\usepackage[utf8]{inputenc} 
\usepackage[T1]{fontenc}    
\usepackage{hyperref}       
\usepackage{url}            
\usepackage{amsfonts}       
\usepackage{amssymb}
\usepackage{nicefrac}       
\usepackage[final,nopatch=footnote]{microtype}      
\usepackage{amsmath}
\usepackage{graphicx}
\usepackage{doi}

\usepackage[style=ieee]{biblatex}
\makeatletter
\let\old@footnotetext\@footnotetext
\renewcommand{\@footnotetext}[1]{%
  \begingroup
    \let\@makefnmark\relax
    \old@footnotetext{#1}%
  \endgroup
}
\makeatother

\title{Maximum Information Extraction Via Clustering and Minimization of Shannon Entropy}

\author{
    \href{https://orcid.org/0000-0002-6306-5229}{\includegraphics[scale=0.06]{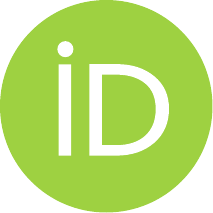}\hspace{1mm}Matteo Becchi}\\
	Department of Applied Science and Technology\\
	Politecnico di Torino\\
	Corso Duca degli Abruzzi, 24, 10129 Torino\\
	\And
	\href{https://orcid.org/0000-0002-3473-8471}{\includegraphics[scale=0.06]{orcid.pdf}\hspace{1mm}Giovanni M. Pavan}\footnote{}\\
	Department of Applied Science and Technology\\
	Politecnico di Torino\\
	Corso Duca degli Abruzzi, 24, 10129 Torino\\
}

\renewcommand{\headeright}{preprint}
\renewcommand{\undertitle}{preprint}
\renewcommand{\shorttitle}{Maximum Information Extraction From Noisy Data}

\hypersetup{
pdftitle={Information gain},
pdfsubject={Maximum Information Extraction Via Clustering and Minimization of Shannon Entropy},
pdfauthor={Matteo Becchi, Giovanni M. Pavan},
pdfkeywords={Information gain, Data clustering, Shannon entropy, Times-series, Data science},
}

\begin{document}

\twocolumn[{%
  \begin{@twocolumnfalse}
    \maketitle

    \begin{abstract}
    In the analysis of any type of system, granting maximum information extraction from its data is non-trivial. Confidence in successful information extraction typically builds on prior knowledge of the studied system or on the user’s experience. However, a robust and objective criterion for ensuring maximum information extraction from data is difficult to define. Here, we introduce a data-driven approach that employs Shannon entropy as a transferable metric to assess and quantify Maximum Information Extraction (MInE) from data via their clustering into statistically-relevant micro-domains. The method is general and can be applied virtually to any type of data or system. We demonstrate its efficiency by analyzing, as a first example, time-series data extracted from molecular dynamics simulations of water and ice coexisting at the solid/liquid transition temperature. The method allows quantifying the information contained in the data distributions (time-independent component) and the additional information gain attainable by analyzing data as time-series (i.e., accounting for the information contained in data time-correlations). The different micro-domains that can be effectively resolved and classified in the system are characterized by own entropy, which are found consistent with experimentally known thermodynamic parameters. A second test case demonstrates how the MInE approach is also effective for high-dimensional datasets and clearly shows how including little informative, but noisy, extra components/features in high-dimensional analyses may be not only useless, but even detrimental to maximum information extraction. This provides a robust parameter-free approach and quantitative metrics for data-analysis, and for the study of any type of system from its data.
    \end{abstract}
    \vspace{10pt}

    \vspace{15pt}

  \end{@twocolumnfalse}
}]

\makeatletter
\@footnotetext{$^*$Corresponding author: \texttt{giovanni.pavan@polito.it}}
\makeatother

Scientific data analysis often requires critical methodological choices -- such as the selection of observables, resolution, and analysis parameters -- that significantly influence both the extraction and interpretation of physical information~\cite{liepe2014framework, bonaccorso2018machine, ding2018model}. These choices are frequently guided by heuristic conventions, prior experience, or implicit assumptions, which can introduce bias and compromise the robustness and reliability of the results~\cite{sweeney2015combined, thrun2021distance}. An automatic method capable of learning directly from the data the maximum information effectively extractable from them would be a breakthrough in many fields, from data science to the study of the physical behavior of complex systems.
As notable examples, various algorithms have been developed to optimize specific analysis parameters~\cite{kohjitani2022gradient, bischl2023hyperparameter}. While effective, these methods typically focus on isolated steps in data acquisition or processing -- such as, e.g., feature selection~\cite{siedlecki1988automatic, arauzo2008consistency, glielmo2022ranking, wild2024maximally}, clustering~\cite{sugiyama2011information, aldana2015clustering}, or coarse-graining~\cite{gkeka2020machine, joshi2021review} -- making it difficult to assess and compare the impact of different choices using a unified transferable metric. However, understanding the physics of complex systems or extracting the maximum amount of information from their data would require a unified framework/method that allows systematically tackling such problems in a more general, holistic, and comprehensive way. 

Here we present a robust, data-driven approach based on information theory to optimize such methodological choices, thereby maximizing the extraction of relevant information through the clustering of data into statistically relevant micro-domains. This method, herein referred to as ``Maximum Information Extraction" (MInE), is versatile and applicable to both uni- and multivariate datasets and to study the physics of virtually any type of system from its data. 
The MInE approach employs a Shannon entropy-based metric to quantify the information gain (i.e., entropy decrease) achieved via clustering across various cases and setups, providing a principled criterion for identifying the optimal methodological choices and analysis setup for maximizing information extraction. 
We demonstrate the effectiveness of MInE in maximizing the extraction of information from data using different case studies. In particular, we show how the method is particularly efficient and useful in cases that are typically non-trivial, such as in noisy datasets and time-series. 

By using Shannon entropy as a transferable metric, MInE compares the information extractable as a function of the micro-clusters that can be effectively resolved/discriminated. The optimal resolution~\cite{doria2025data} and best analysis setup for maximum information extraction are identified as those minimizing Shannon entropy. 
We demonstrate the effectiveness of MInE through two case studies. First, we compare the information extractable from univariate time-series data -- obtained using different descriptors~\cite{bartok2013representing, caruso2023timesoap, crippa2023detecting} from Molecular Dynamics (MD) trajectories of water and ice phases coexisting at the melting temperature -- as a function of resolution and/or descriptor choice. We demonstrate how the different micro-clusters that are discovered from the data correspond to domains that are characterized by their own internal entropy -- proving their physical relevance --, whose differences are found consistent with known thermodynamic quantities for aqueous systems (e.g., entropy of fusion). Second, in a model Langevin dynamics on a bi-dimensional energy landscape, we show how MInE can assess how information extraction is influenced by the interplay between the number of dimensions considered and the impact of noisy data and frustrated information phenomena~\cite{lionello2025relevant} in multivariate analyses. These case studies highlight MInE as a physically robust and general method for optimizing data analysis and for maximizing the knowledge attainable from virtually any type of system and/or data.  

\section*{Theoretical Framework}

The Shannon entropy~\cite{shannon1948mathematical} of an observable $x$, which takes values in a set $\mathcal{X}$ with probability distribution $p(x)$, is defined as: 

\begin{equation}
    H(x) = -\sum_{x\in\mathcal{X}}p(x)\log_2 p(x)
    \label{eq:shannon}
\end{equation}

This quantity measures the uncertainty in $x$, with higher entropy indicating a broader distribution and lower entropy corresponding to more sharply concentrated distributions. By normalizing the Shannon entropy between 0 and 1 (see Methods section), we can define the information available from the data as: 

\begin{equation}
    I = 1 - H
    \label{eq:infotot}
\end{equation}

In this way, the information contained in the data takes values in the range $0\leq I \leq 1$. In the limit case where the data contain no uncertainty (e.g., all data points exactly equal to one specific single value), $I$ equals to 1, while $I=0$ corresponds to the case where data are completely random (pure entropy), containing no structure, and no statistically relevant information can be extracted from them. 

In practice, the information $I_0$ contained in the data can be increased by clustering them into statistically distinct micro-domains. For a dataset clustered into $K$ clusters with probability distributions $p_k(x)$, the Shannon entropy of the clustered system is:

\begin{equation}
    H_{\mbox{clust}}(x) = \sum_{k=1}^K f_k H_k
    \label{eq:h_clust}
\end{equation}
where $f_k$ is the fraction of data points in cluster $k$, and $H_k$ is the Shannon entropy of the data within that cluster. By convexity, clustering never increases entropy, and the corresponding information gain is thus defined as: 

\begin{equation}
    \Delta I = -\Delta H = H_0 - H_{\mbox{clust}} \geq 0
    \label{eq:info_gain}
\end{equation}
where $H_0$ is the Shannon entropy of the original data distribution. 

This quantity effectively measures the reduction in uncertainty achieved via data clustering~\cite{soofi2010information}. Based on Eq.~\ref{eq:info_gain}, $I_\text{clust} = I_0 + \Delta I$. Optimizing the clustering of the data thus allows to maximize the information gain following clustering, and consequently the information effectively extractable from the data, $I_\text{clust}$. 
A trivial clustering -- where all points belong to a single cluster, or are assigned randomly -- yields $\Delta I \sim 0$, indicating little to no information gain compared to that attainable simply form the distribution of the data. In contrast, an effective clustering maximizes $\Delta I$, maximizing the translation of the data structure into information. 

Detailed information on the entropy calculation are provided in the Materials and Methods section, as well as in the SI.
Note that, while here Shannon entropy is our primary measure for building our method, alternative information-theory based metrics (such as, when dealing with time-series, Approximate Entropy~\cite{pincus1991approximate} or Sample Entropy~\cite{richman2000physiological, richman2004sample}) can also be employed within the same framework. 

In short, MInE shows the maximum amount of information attainable from data, and unveils how to effectively extract it. Note that the MInE workflow thus substantially differs from entropy-minimization based clustering algorithms~\cite{palubinskas1998unsupervised}. Shannon entropy minimization is not used in MInE to guide the clustering itself, but it is rather applied a posteriori to evaluate the effectiveness of the methodological choices for the clustering and to quantify the information attainable from it. 
While the MInE approach is general and can be used to reveal the maximum extractable information from virtually any type of data (e.g., static, dynamic), here we demonstrate its effectiveness by applying it to prototypical challenging cases, such as highly noisy time-series data generated by complex dynamical systems of various types. 

\section*{Results and Discussion}

\subsection*{From entropy to information and back}

\begin{figure}[htbp]
  \centering
  \includegraphics[width=\linewidth]{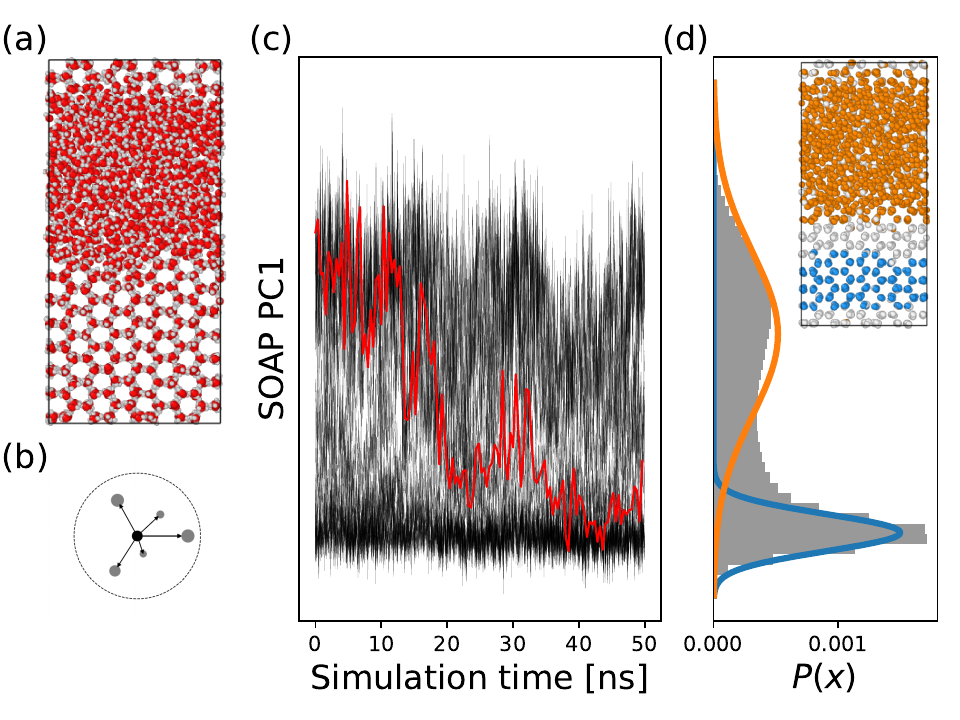}
  \caption{(a) Snapshot of the water/ice atomistic model system used as a case study. (b) Schematic of the SOAP descriptor, which provides a high-dimensional representation of the density, order/disorder, and symmetry of the arrangement of neighboring molecules around each molecule. (c) Denoised~\cite{donkor2024beyond} PC1 SOAP time-series~\cite{lionello2025relevant, martino2024data} for the 2048 molecules as a function of simulation time. In red: the signal of a representative molecule undergoing a water-to-ice transition. (d) Probability distribution $P(x)$ of the SOAP PC1 signals (gray). Two main clusters, corresponding to the liquid and ice phases (inset), are identified by the two maxima in $P(x)$ (orange and blue Gaussian fits). These clusters are readily detected using pattern recognition methods (see SI for details). }
  \label{fig:fig1}
\end{figure}

\begin{figure*}[htbp]
  \centering
  \includegraphics[width=\linewidth]{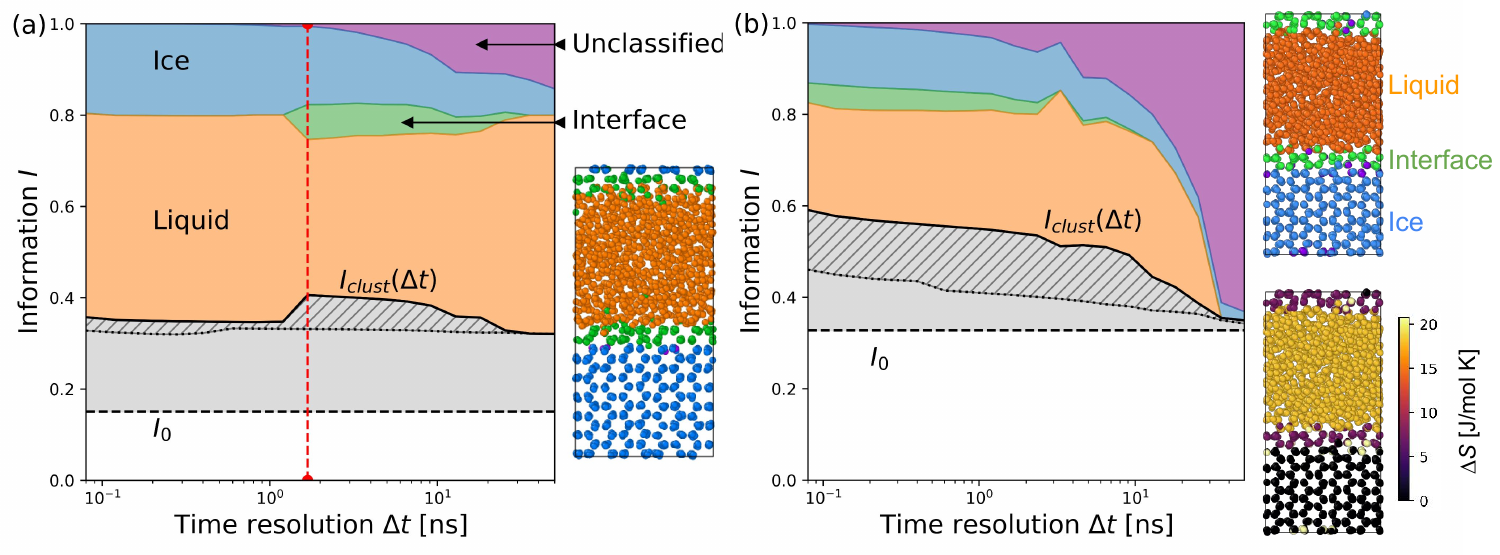}
  \caption{Information gain from Onion clustering as a function of the time resolution $\Delta t$ used in the analysis. 
  Panel (a) shows results for the SOAP PC1 dataset with spatial averaging; panel (b) shows results for the LENS dataset. In both panels, the horizontal dashed line ($I_0$) represents the information content of the raw data distribution. The solid black curve ($I_\text{clust}$) represents the information retained after clustering, which varies with $\Delta t$. The gray-shaded area corresponds to the increase in information content achieved through clustering. The dotted line shows the information gain obtained when the dataset frames are randomly reshuffled, while the diagonally hatched area highlights the additional gain attributable to time correlations. 
  The width of the colored regions reflects the weighted Shannon entropy $f_k H_k$ of each environment, illustrating how entropy varies with $\Delta t$. 
  In panel (a), $I_\text{clust}$ reaches a maximum at $\Delta t \sim 2$~ns (vertical red dashed line), indicating the optimal time resolution for extracting information from the SOAP PC1 time-series. The simulation snapshot is colored according to Onion clustering at this resolution: bulk ice and liquid water appear in blue and orange, respectively, with the solid–liquid interface in green. A small fraction of unclassifiable points appear as sparse molecules in purple (color coding matches that in the plot). 
  On the right: The top snapshot shows the corresponding clustering, using the same color scheme as in panel (a). The bottom snapshot is the same one, colored by entropy difference calculated relative to that of the bulk of ice and converted in units of [J~mol$^{-1}$~K$^{-1}$]. }
  \label{fig:fig2}
\end{figure*}

As a first example, we tested the MInE framework to analyze MD simulation trajectories of 2048 TIP4P/ICE~\cite{abascal2005potential} molecules coexisting in dynamic equilibrium between solid and liquid phases at the melting temperature (Fig.~\ref{fig:fig1}a). After equilibration, we performed a $\tau=50$~ns production run under $NPT$ conditions, sampling trajectories every 0.04~ns (complete simulation details are provided in the SI). 
We used various single-particle descriptors to extract univariate time-series data from the MD trajectories, each capturing distinct structural and/or dynamical features of the system, with its own signal-to-noise ratio~\cite{donkor2023machine, martino2024data}. For example, Figs.~\ref{fig:fig1} and~\ref{fig:fig2} show results obtained using the Smooth Overlap of Atomic Positions (SOAP)~\cite{bartok2013representing},  which provides a rotationally invariant, high-dimensional representation of the local molecular density around each molecule (Fig.~\ref{fig:fig1}b). The SOAP power spectrum offers a fingerprint of the spatial arrangement of neighboring molecules within a sphere of radius $r_c$ around each molecule. Fig.~\ref{fig:fig2} also contains analyses on the Local Environment and Neighbors Shuffling (LENS) descriptor~\cite{crippa2023detecting}, which captures the entity and speed of the reshuffling of the neighboring molecules of each molecule. 

In the case of noisy time-series data, such as these ones, the ability to detect/discriminate statistically relevant micro-domains through single point clustering depends on the resolution $\Delta t$ used to segment and analyze the data~\cite{doria2025data, becchi2024layer, lionello2025relevant}. Among the available algorithms for single point time-series analysis~\cite{aghabozorgi2015time}, here we use Onion clustering~\cite{becchi2024layer}, an unsupervised method that identifies all statistically relevant micro-domains (including hidden ones) that can be resolved in a time-series of length $\tau$ as a function of $\Delta t$, as well as the fraction of unclassifiable data due to insufficient resolution. More details about Onion clustering are available in the SI. 
While Onion clustering is particularly well suited to for handling noisy time-series data, and it is thus chosen as a basis for the demonstrations presented herein, note that the MInE method is general and can be applied, in principle, in combination with any other clustering technique. 

In the following, we show as a representative example the results obtained by analyzing denoised~\cite{donkor2024beyond} time-series data of the first principal component (PC1) of the SOAP vectors, which is an interesting informative descriptor for such systems~\cite{lionello2025relevant, martino2024data} and for the demonstrations that we are providing herein (complete description for the SOAP analysis are provided in the Supplementary Material). 
Data obtained with other descriptors~\cite{crippa2023detecting} and/or by reducing the dimensionality of the SOAP spectra with other methods (e.g., via Time-lagged Independent Component Analysis~\cite{schwantes2015modeling, hoffmann2021deeptime}) are reported later in the paper, and in the Supplementary Material. 
Fig.~\ref{fig:fig1}c shows the SOAP PC1 time-series data for all molecules along the trajectory (in black), with one molecule undergoing an ice-to-water transition highlighted in red. The probability distribution $P(x)$ of the entire SOAP PC1 dataset is shown in Fig.~\ref{fig:fig1}d (gray). The two prominent density peaks in $P(x)$ allow typical clustering methods to readily identify two main clusters in the data. These clusters, represented by the orange and blue Gaussian curves centered on the $P(x)$ peaks in Fig.~\ref{fig:fig1}d, correspond to the liquid and solid phases, as shown in the snapshot inset with matching color coding. 

However, additional information is nested inside these data, particularly in their temporal correlations~\cite{becchi2024layer}. This is evident from the results obtained by Onion clustering at smaller $\Delta t$ (i.e., increasing the temporal resolution). Complete Onion clustering results are shown in Fig.~S1 in the SI, showing that the number of resolvable clusters is maximized in the resolution interval 2~ns $\lesssim \Delta t \lesssim$~20~ns. Within this range, the method identifies three statistically distinct clusters corresponding to the ice and water phases, and the ice/water interface (see Fig.~S1 in the SI). 

Note that results such as, e.g., the number and quality of detected clusters, as well as the fraction of unclassified data, are descriptor-dependent. Each descriptor is characterized by its own signal-to-noise ratio, and its feature space is defined by metrics that differ from those of other descriptors. As a result, it is not trivial to unambiguously infer, for example, whether the information captured by the SOAP PC1 descriptor represents the maximum extractable information, or how one descriptor compares to another. MInE overcomes this limitation by using entropy as a transferable metric and exploiting Shannon entropy minimization to assess the maximum information that can be extracted from the data following clustering, as defined by Eq.~\ref{eq:info_gain}. 

Fig.~\ref{fig:fig2}a shows the information $I$ attainable from the data before and after clustering. The information contained in the data probability distribution $P(x)$ (prior to clustering) can be quantified by its Shannon entropy, $H_0$ (calculated via Eq.~\ref{eq:shannon}), as $I_0 = 1 - H_0\sim 0.151$ (black horizontal dashed line in Fig.~\ref{fig:fig2}a). Since this value depends solely on the data distribution, it is independent of the time resolution $\Delta t$ used in the subsequent clustering analysis. 
In contrast, the information content after clustering is given by $I_\text{clust}(\Delta t) = 1 - H_\text{clust}(\Delta t)$, represented by the solid black curve in Fig.~\ref{fig:fig2}a. The gray-shaded area between these two curves represents the total information gain extractable, in this case, by Onion clustering. Shannon entropy minimization is achieved at $\Delta t^* =1.8$~ns (vertical red dashed line), where the maximum information $I_\text{max} = I_\text{clust}(\Delta t^*) \sim 0.41$ is attained. 
At all resolutions, the sum of the resolved information $I$ and the residual entropy $H$ remains constant (by definition). Above the solid black line in Fig.~\ref{fig:fig2}a, the weighted Shannon entropy of the different resolved environments (micro-clusters), $f_k H_k$, is shown for each value of $\Delta t$. As seen in Fig.~\ref{fig:fig2}a, within the range of $\Delta t$ where the ice/water interface (green area)is resolved as a distinct micro-cluster, the information $I_\text{clust}$ is maximized when the fraction of unclassified data points (in violet) is minimized. 

From an information-theoretic perspective, the disappearance of the interface cluster (green) leads to an increase in the residual entropy $f_k H_k$ of the remaining clusters (Fig.~\ref{fig:fig2}a). Conversely, when the interface is detected, the total entropy in Eq.~\ref{eq:h_clust} decreases, despite the introduction of an additional term. This is because the reduction in $f_k H_k$ for the ice and liquid phases more than compensates for the added entropy contribution of the interface. 
Physically, the optimal range 2~ns~$\lesssim \Delta t \lesssim$~20~ns provides the best balance between resolution and noise, enabling the identification of the ice/water interface, which has a characteristic residence time of $\sim 10$~ns~\cite{karim1988ice}. 
At lower time resolutions ($\Delta t \geq 20$ ns), an increasing fraction of data points remain unclassified, resulting in a raise in the residual entropy $H_\text{clust}$ and a corresponding decrease in the information gain (gray area in Fig.~\ref{fig:fig2}a). 
On the other hand, at higher resolutions ($\Delta t \leq 1$ ns), the interface can no longer be resolved as a distinct environment. This occurs because excessively fine temporal segmentation emphasizes short-time molecular vibrations --interpreted as noise-- at the expense of capturing the physically meaningful transitions between phases~\cite{lionello2025relevant}. 

To estimate the contribution of time correlations to the total information $I_\text{clust}$, we repeated the Onion clustering analysis after randomly reshuffling the time-series frames. This procedure preserves the data probability distribution $P(x)$, while effectively removing temporal correlations, thereby eliminating any information that could be extracted from them. 
We performed this reshuffling ten times and computed the average values of $I_\text{clust}$ as a function of $\Delta t$; the variance in the estimation was consistently below $0.1\%$ for all $\Delta t$. The results are shown in Fig.~\ref{fig:fig2}a as a dotted curve. Notably, this curve is nearly horizontal, indicating that -- once time correlations are removed -- the extractable information becomes largely independent of the resolution $\Delta t$. Furthermore, the value of $I_\text{clust}$ obtained from the reshuffled data closely matches that obtained from the original time-series when analyzed at the lowest possible resolution ($\Delta t = \tau = 50$~ns). In Fig.~\ref{fig:fig2}a, the difference between the $I_0$ baseline (dashed line) and the reshuffled $I_\text{clust}$ (dotted line) represents the information gain achievable via clustering under a purely ergodic approximation -- that is, neglecting time correlations. 
Finally, the difference between this dotted $I_\text{clust}$ curve and the solid black $I_\text{clust}(\Delta t)$ curve (corresponding to the original time-ordered series) quantifies the additional information that is uniquely encoded in the temporal correlations -- highlighted by the diagonally hatched gray area in Fig.~\ref{fig:fig2}a. 

Fig.~\ref{fig:fig2}b shows the same analysis on time-series data obtained from the same MD trajectories, but using a different single-particle descriptor, LENS~\cite{crippa2023detecting}. LENS -- which generally exhibits a higher intrinsic signal-to-noise ratio compared to SOAP~\cite{doria2025data, martino2024data, becchi2024layer, caruso2025classification} -- enables correct detection of the interface from the highest time resolution, $\Delta t = 0.08$~ns, up to approximately $\Delta t\sim10$~ns. 

To further assess the physical relevance of the Shannon entropy values computed in this analysis, we attempted to relate them to the thermodynamic entropy of the system under study. This comparison can be made following, for instance, the approach presented in Ref.~\cite{hong2025generalized}, under the assumption that the chosen descriptor captures all of the system's degrees of freedom and their correlations. While this assumption may not strictly hold -- since any descriptor is, to some extent, incomplete -- we show below that, for the LENS data analyzed here, the results obtained are reasonable and physically meaningful. 

Absolute entropy values computed on continuous variables, such as LENS in our case, are generally not meaningful, as they depend strongly on the binning used in the Shannon entropy estimation. However, entropy differences can still carry significant physical meaning. 
Using the LENS signals we compute the entropy difference between the solid ice and liquid water micro-domains, obtaining a value of $\Delta S\sim18$~J~mol$^{-1}$~K$^{-1}$. This is very close to the experimental entropy of fusion of water, $\Delta S_\text{melt} \sim 22$~J~mol$^{-1}$~K$^{-1}$~\cite{lide1995crc} (especially considering the accuracy that can be expected from such simulations, and the description allowed by the LENS descriptor). 
This demonstrates the robustness of the approach and the physical significance of the results obtained, in a purely data-driven way, using the MInE method. 

Noteworthy, this approach also allows us to estimate the entropy difference between any other micro-state in the system -- in this case, the bulk solid ice vs. the solid–liquid interface environments, a quantity that is generally difficult to determine experimentally or otherwise. For the ice/water interface, we obtain a value of $\Delta S_\text{interf}\sim7.2$~J~mol$^{-1}$~K$^{-1}$ compared to the bulk of ice. The lower snapshot in Fig.~\ref{fig:fig2}b is colored according to the entropy difference between each environment vs. solid ice. 
This is a significant result, as it clearly highlights the physical relevance of the interface environment in such a system. The fact that this micro-cluster corresponds to a dynamically distinct molecular environment, with its own entropy -- different from those of both ice and liquid water --, implies that it should not be confused with either of these two other phases. 
In particular, any method or analysis that fails to identify the interface environment as a separate and distinct cluster (i) loses important physical information and (ii) compromises the statistical characterization of the ice and liquid domains (especially when interface water molecules are incorrectly assigned to one of the two bulk phases~\cite{crippa2023machine}). 

Note that other tested descriptors, which account for different degrees of freedom, yield different entropy values (although they all remain within the same order of magnitude). For example, using SOAP PC1, we obtain a ice-water $\Delta S\sim10.7$~J~mol$^{-1}$~K$^{-1}$. This difference between LENS and SOAP can be rationalized by considering that these descriptors capture only part of the system's internal degrees of freedom. In particular, these results demonstrate how the difference between the solid ice and liquid water phases is better captured by differences in the local diffusivity of the molecules populating these two environments (measured by LENS) rather than by differences in their local structural vibrations (measured by SOAP). 

\subsection*{The information captured by different descriptors}

\begin{figure*}[htbp]
  \centering
  \includegraphics[width=\linewidth]{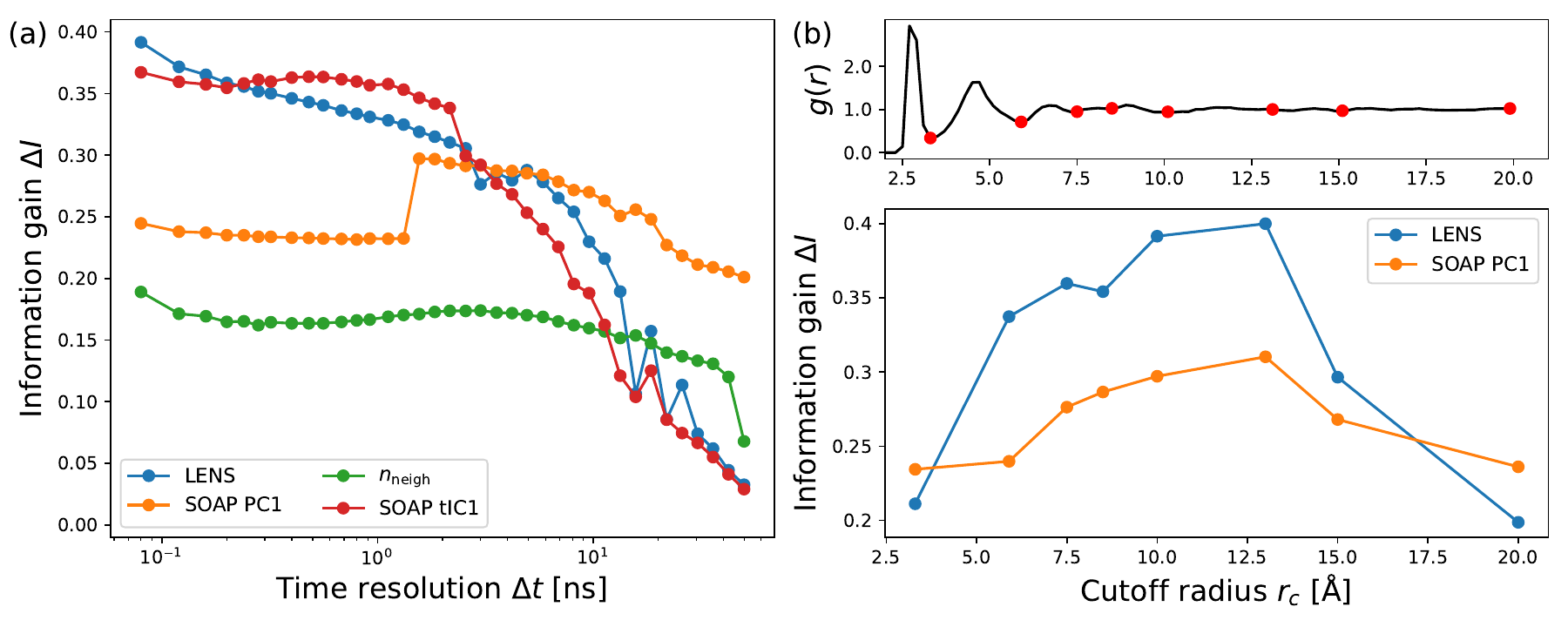}
  \caption{Optimizing descriptor choice and spatiotemporal resolutions for maximum information extraction. 
  (a) Relative information gain $\Delta I/H_0$, obtained by applying Onion clustering to time-series data derived from different descriptors:  LENS (blue), the first principal component (PC1) of SOAP (orange), the first time-lagged independent component (tIC1) of SOAP (red), and the number of neighbors $n_\text{neigh}$ (green). 
  (b) Top: Radial distribution function $g(r)$ of the water molecules (oxygen atoms only), computed over the entire simulation trajectory. Red dots mark the minima in $g(r)$, corresponding to solvation shells, used here as characteristic cutoff values $r_c$ for computing descriptor time-series. Bottom: Maximum relative information gain $\Delta I/H_0$ attainable via Onion clustering on LENS and SOAP PC1 time-series calculated using different cutoff radii $r_c$. }
  \label{fig:fig3}
\end{figure*}

By assessing the maximum amount of information that can be extracted from a given dataset, MInE also enables a direct comparison, and ranking, of different descriptors based on their effectiveness in capturing relevant information from the same trajectories. This comparison is made possible by the transferable nature of the Shannon entropy–based metric used. 
As shown in Fig.~\ref{fig:fig3}a, we applied MInE to time-series data derived from different descriptors computed on the same MD trajectories of the ice/water model system. In addition, we employed alternative dimensionality-reduction techniques to analyze the high-dimensional SOAP spectra. 

For this specific system, the results of these analyses, shown in Fig.~\ref{fig:fig3}a, reveal that some descriptors enable the conversion of MD trajectories into time-series from which significantly more information can be extracted than others. In particular, these descriptors yield higher relative information gain $\Delta I/H_0$. 
For example, the LENS descriptor~\cite{crippa2023detecting} and the first time-lagged Independent Component (tIC1)~\cite{schwantes2015modeling, hoffmann2021deeptime} of the SOAP spectra, when analyzed using Onion clustering, provide the highest information gain (up to time-resolutions of $\Delta t \leq 2$ ns) among the descriptors tested: $\Delta I/H_0 \sim 0.35-0.4$. This relative information gain highlights the effectiveness of these descriptors in retaining and extracting physically relevant information encoded in the molecular trajectories. 
Note that all four descriptors used in this comparative analysis were computed with a cutoff radius of $r_c=10$~\AA, corresponding to the third solvation shell. This parameter sets the inter-particle distance over which correlations, reconfigurations, and structural features are probed. As such, $r_c$ is a fundamental parameter that determines the spatial resolution of the analysis and directly influences the amount and type of information that can be effectively captured. 

\subsection*{Optimal resolution for maximum information extraction}

Many widely used single-particle descriptors are designed to capture the behavior of neighboring units surrounding each particle in the system. The neighborhood of each molecule is typically defined as a sphere with a cutoff radius $r_c$, which sets the spatial scale over which correlations, rearrangements, symmetries, and other structural features are analyzed. In this way, $r_c$ effectively determines the spatial resolution of the analysis and, consequently, the physical information that can be extracted from the system. 
Recent work has shown that the optimal spatial resolution for studying a given system depends on whether the dominant physical phenomena are local or collective in nature~\cite{doria2025data}. The MInE approach offers a quantitative framework to identify this optimal resolution by evaluating how much relevant information can be extracted at different values of $r_c$. 

As a proof of concept, we tested this approach on the same MD trajectory of the TIP4P/ice solid–liquid coexistence system. We computed both LENS and SOAP descriptors using different cutoff radii, $r_c$, chosen to correspond to characteristic minima in the water–water radial distribution function, $g(r)$ (see Fig.~\ref{fig:fig3}b, top; red dots). 
Fig.~\ref{fig:fig3}b (bottom) shows how the maximum relative information gain, $\Delta I/H_0$, achievable through Onion clustering varies as a function of the cutoff radius used to compute the descriptor time-series. For both descriptors, $\Delta I/H_0$ exhibits a clear maximum at $r_c \sim 10$–13~\AA, thereby identifying this as the optimal resolution for maximizing information extraction in this system. 

As discussed in detail in Ref.~\cite{doria2025data}, this result reflects the characteristic length scale of the collective structural dynamics dominating aqueous systems of this kind: accurately capturing such phenomena requires including at least up to the third solvation shell. Smaller cutoff values fail to capture essential mid-range correlations, resulting in datasets dominated by local noise and fast vibrations; conversely, excessively large cutoffs cause information loss due to over-averaging. 
More broadly, the results shown in Fig.~\ref{fig:fig3} illustrate how MInE offers a robust, quantitative framework for optimizing key analysis parameters -- such as descriptor type, cutoff radius, and spatial and temporal resolutions -- to ensure maximal information extraction from molecular trajectory data. 

\subsection*{Application to multivariate datasets}

\begin{figure}[htbp]
  \centering
  \includegraphics[width=\linewidth]{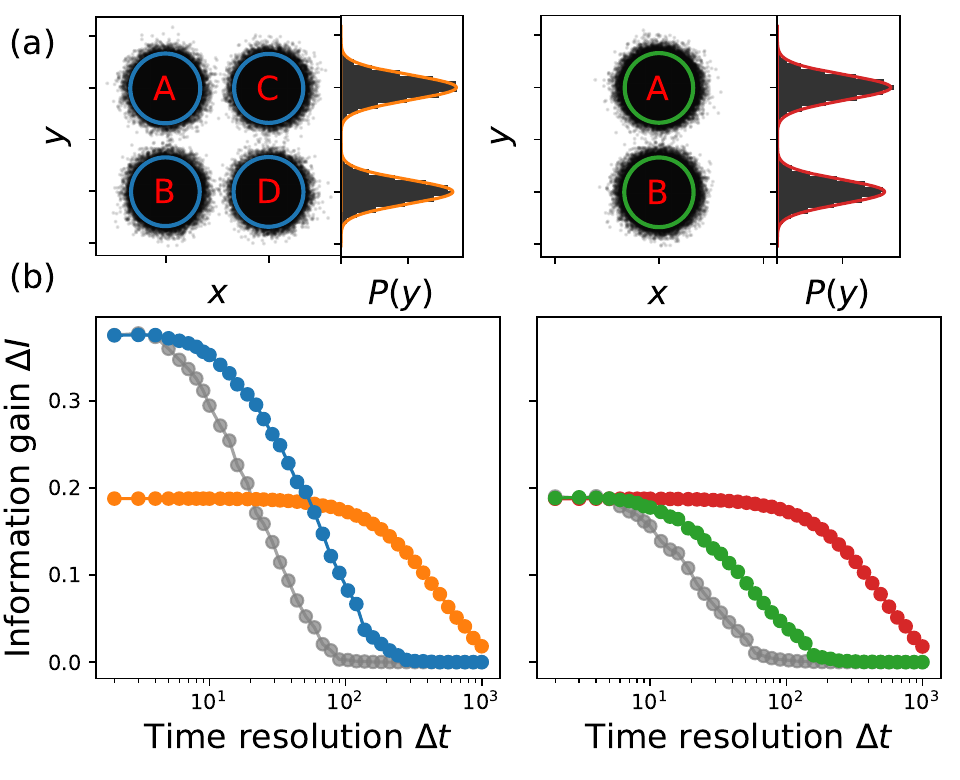}
  \caption{(a) From left to right, the trajectories of 100 particles in a bi-dimensional energy landscape with 4 and 2 minima respectively, and the probability distribution of the $y$ coordinate $P(y)$. The clusters identified by Onion clustering in the two systems, using the full $(x, y)$ trajectories or only the $y$ coordinate, are shown in blue, orange, green and red respectively. 
  (b) Information gain $\Delta I$ from Onion clustering as a function of time resolution $\Delta t$ (in simulation time-steps) used for the analysis. For each of the two example model systems (left and right panels), three different datasets are analyzed: different color curves matching those in panel (a). For the system with four minima (left), clustering with $(x,y)$ (blue curve) allows resolving all four A-D minima, yielding at maximum twice the information gain compared to that attainable when clustering using variable $y$ only (orange: which allows resolving only two minima), but the performance degrades an order of magnitude earlier ($\Delta t\sim10$ vs. $100$ frames). For the two-minima system (right), clustering with $(x,y)$ (green) and with $y$ (red) achieves the same maximum gain (allowing to resolve in both cases both A-B minima), though degradation is again faster for the bi-variate case. In both systems, we added a third case (gray curve) demonstrating how adding a third $z$ coordinate, which does not bring additional relevant information but just noise (the minima become spherical in three dimension), does not increase the attainable information gain but also accelerates analysis degradation confining it to lower $\Delta t$. }
  \label{fig:fig4}
\end{figure}

MInE can also be applied to high-dimensional datasets and multivariate time-series. When using multivariate descriptors, quantifying information becomes particularly relevant in tasks such as feature selection and dimensionality reduction. These approaches aim at reduce data complexity but often involve trade-offs, as decreasing the number of variables can lead to information loss. To explore this, we tested the MInE framework on a simple multivariate model dataset. Specifically, we simulated the Langevin dynamics of 100 particles in two distinct bi-dimensional potential energy landscapes: one with four minima (A to D), requiring both $(x, y)$ coordinates for proper identification, and another with only two minima (A and B), distinguishable using the $y$ coordinate alone. In both cases, all minima are subject to identical Gaussian noise (Fig.~\ref{fig:fig4}a). 
We then applied the same MInE procedure as described above to four different datasets, obtained by analyzing both systems using either the full $(x, y)$ coordinates or only the $y$ coordinate through Onion clustering (see SI for full details on these test cases). 

The information gain obtained after clustering for these model systems is shown in Fig.~\ref{fig:fig4}b. For the system with four energy minima (Fig.~\ref{fig:fig4}b, left), using both $(x, y)$ coordinates all four minima to be identified. In contrast, using only the $y$ coordinate does not distinguishes between A and C or between B and D, which collapse into two degenerate (and doubly-populated) A+C and B+D states. 
Comparing the blue and orange curves highlights that the maximum attainable information gain (reached in this system for $\Delta t< 5$ frames) is twice as high when performing a bi-dimensional analysis (blue) compared to a mono-dimensional one (orange). 
For the system with just two minima (Fig.~\ref{fig:fig4}b, right), the maximum attainable information gain is the same wether using both $(x, y)$ or just the $y$ coordinate (green vs. red curves): in this case, both minima are correctly identified regardless of dimensionality, making the second variable unnecessary. 
Importantly, the maximum information gain is identical in the orange curve (mono-dimensional analysis of the four-minima system) and the red and green curves (two-minima system), despite representing different systems and analyses. This demonstrates how Shannon entropy within the MInE framework serves as an objective, transferable, and quantitative metric for assessing and comparing the extractable information across datasets and analysis strategies. 

Interestingly, in both systems, the information gain (and thus clustering performance) decreases more rapidly with increasing $\Delta t$ in the bi-dimensional analyses (blue and green curves) compared to the mono-dimensional ones (orange and red). 
In the four-minima system, one initial hypothesis attributed this behavior to the ``apparent'' longer residence times of particles in the degenerate A+C and B+D states identified by the mono-dimensional analysis, compared to the shorter residence times in the individual A, B, C, and D states resolved by the bi-dimensional one. 
However, this explanation does not hold for the two-minima system (Fig.~\ref{fig:fig4}b, right), where both A and B states are correctly identified in both types of analyses -- yet the same faster decrease in information gain with increasing $\Delta t$ is observed in the bi-dimensional case. 
These findings for the two-minima case point instead to the role of noise-addition phenomena that can affect high-dimensional analyses: specifically, even though the $x$ component in this case does not provide relevant information, it still introduces noise~\cite{lionello2025relevant}. 
In higher-dimensional spaces, clustering algorithms tend to leave a larger fraction of data points unclassified~\cite{assent2012clustering, lionello2025relevant}, as also shown in Fig.~S2 of the SI. 
As discussed elsewhere~\cite{lionello2025relevant}, these results underscore how including additional (noisy) dimensions may be not only unhelpful, but actually detrimental to the effectiveness of high-dimensional analyses. 

To further validate these effects, we performed an additional test in both systems by introducing a third dimension $z$, which -- like $x$ in the previous test -- does not carry any meaningful information but only contributes Gaussian noise (i.e., all minima are characterized by identical, uncorrelated noise in all directions). 
The gray curves in Fig.~\ref{fig:fig4}b,c show that while the maximum attainable information remains unchanged compared to the lower-dimensional cases, the range of $\Delta t$ values over which this information is effectively extractable becomes further reduced. 
This confirms that adding increasingly noisy components hinders the extraction of meaningful information, particularly that embedded in the time-correlations of the data. 
These findings reflect a manifestation of the so-called ``curse of dimensionality''~\cite{kouiroukidis2011effects, altman2018curse}, highlighting that in clustering analyses there exists a threshold beyond which the inclusion of additional dimensions shifts from enriching the information content to merely introducing noise. MInE proves particularly valuable in this context, as it enables probing and quantifying these effects in a rigorous and interpretable way. 

We underline that, while the results above focus on time-series data -- where temporal correlations play a key role -- the MInE approach is general and applicable to other types of datasets as well. 
For example, in the case of static data distributions (or in situations where time correlations are irrelevant or difficult to resolve), the $I_\text{clust}$ becomes independent of the $\Delta t$.
In such contexts, MInE can be thus used to identify, e.g., the most effective clustering method, or parameter set (descriptor(s)) that allows minimizing Shannon entropy and maximizing information extraction from the data distributions. 

\section*{Conclusions}

We introduce a general data-driven framework to quantify the information extractable from virtually any type of data and to assess Maximum Information Extraction (MInE). The method relies on Shannon entropy minimization to quantify how much structure in the data can be effectively resolved via data clustering. In a purely unbiased and data-driven way, MInE allows to optimize data analysis (e.g., the choice of the best descriptor, or of the optimal spatial and temporal resolutions) to maximize the extraction of physically meaningful information. The framework is applicable to both continuous and categorical data and provides a transferable metric to guide analysis across diverse systems. 

Here, we first test the efficiency of the MInE approach to analyze molecular dynamics simulations of solid–liquid water coexistence. Single point time-series clustering conducted by Onion clustering allows maximizing the Shannon entropy reduction following to clustering. MInE reveals the best suited descriptors (among a set of tested ones) and the optimal resolutions to maximum information extraction. We demonstrate how this maximum information extraction aligns with the robust detection of the solid–liquid interface as a distinct environment (Fig.~\ref{fig:fig2}). 

In one shot, MInE estimates the Shannon entropy of the various detected environments, and it allows to reconduct it to thermodynamic entropy differences. As a notable example, the comparison between the entropy of the bulk ice and liquid water provides an entropy difference between the solid and liquid phases in the system that is in good agreement with the experimentally known thermodynamic entropy of fusion of water (Fig.~\ref{fig:fig2}b). This is a remarkable result, considered the typical accuracy expected from such simulations and the level of description allowed by one single descriptor, i.e., LENS. 
Any individual descriptor is, per se, incomplete to some extent, and different descriptors may provide different entropy difference estimations. This demonstrates the extent to which different representations encode physically relevant degrees of freedom and proves how MInE can be a fundamental tool to compare, assess, and rank descriptors based on their completeness and ability to maximize information extraction (Fig.~\ref{fig:fig3}).

Noteworthy, MInE also allows estimating the entropy of the solid–liquid interface domain relative to bulk ice (Fig.~\ref{fig:fig2}b), a quantity that is difficult to access experimentally.
MInE proves the relevance of this micro-state in the system, demonstrating how this has its own entropy, which is different from those of ice and liquid phases. This demonstrates how MInE can not only assist the discovery of new relevant micro-states in the system, but it also reveals their relative entropy values and quantifies the information gain associated to their discovery. This is useful for validating the obtained results in the case of known systems (as in the aqueous system studied herein, for which experimental entropy variations are available). Also, this is especially useful for systems about which little is known a priori, for which MInE can reveal precious and robust quantitative information on their internal microscopic-level physics. 

The results of Fig.~\ref{fig:fig4} demonstrate the efficiency of MInE in analyzing also high-dimensional dataset, revealing the information loss encountered when, e.g., missing one fundamental information. This provides a robust quantitative handle to support dimensionality reduction approaches as well as feature selection to maximum information extraction. 

The MInE method is simple to implement, physically interpretable, and it does not rely on any system-specific assumptions. This provides a general framework to quantify information gain in data analysis that can be applied to a wide range of problems, from atomistic to biological systems, as well as to other complex datasets. We expect that MInE will become a fundamental framework for studying and resolving the physics of systems from their data, as well as to guide the discovery of new knowledge via data analysis in general. 

\section*{Methods}
\subsection*{Shannon entropy calculation}
In our implementation, entropy $H$ is estimated from the histograms of the dataset before and after clustering, using a consistent binning scheme to ensure comparability. To facilitate interpretation, all entropy values (which following Eq.~\ref{eq:shannon} are measured in bits) in this work are normalized with respect to the maximum possible entropy value $\log_2 n_b$~bit, providing an adimensional quantity (see SI for complete details on the method). 

\subsection*{MD simulations}
Complete details on the MD simulation performed for this study are reported in the SI.

\subsection*{Code implementation and availability}
All the code and data necessary to reproduce the analysis of this work are available on a Zenodo repository~\cite{zenodo} at \href{https://doi.org/10.5281/zenodo.15236523}{https://doi.org/10.5281/zenodo.15236523}.

\section*{Acknowledgments}
The authors thank Chiara Lionello for the insightful discussions. 
G.M.P. acknowledges the support received by the European Research Council (ERC) under the Horizon 2020 research and innovation program (grant agreement no. 818776 - DYNAPOL). 

\printbibliography

\onecolumn
\renewcommand{\thefigure}{S\arabic{figure}}
\renewcommand{\theequation}{S\arabic{equation}}
\setcounter{figure}{0}
\setcounter{equation}{0}

\section*{Supporting Information for ``Maximum Information Extraction Via Clustering and Minimization of Shannon Entropy''}
Matteo Becchi, Giovanni M. Pavan

\subsection*{Supporting Figures}

\begin{figure}[htbp]
    \centering
    \includegraphics[width=0.5\linewidth]{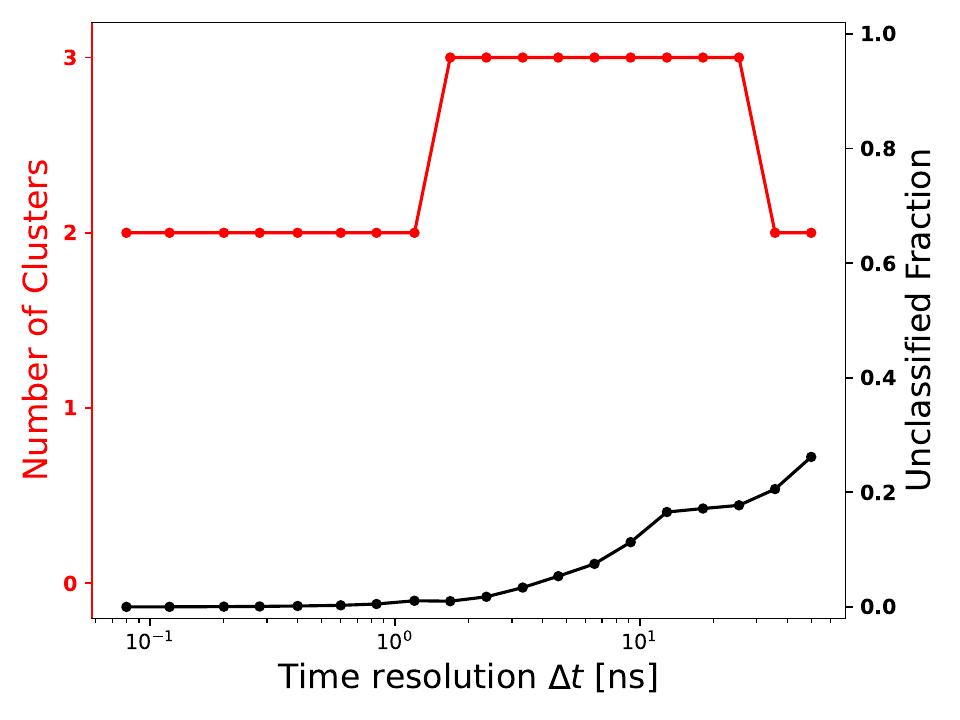}
    \caption{Number of cluster discovered (red) and fraction of unclassified data points (black) as a function of the time resolution $\Delta t$ used for the onion clustering on SOAP PC1 data. }
    \label{fig:figs1}
\end{figure}

\begin{figure}[thbp]
    \centering
    \includegraphics[width=0.5\linewidth]{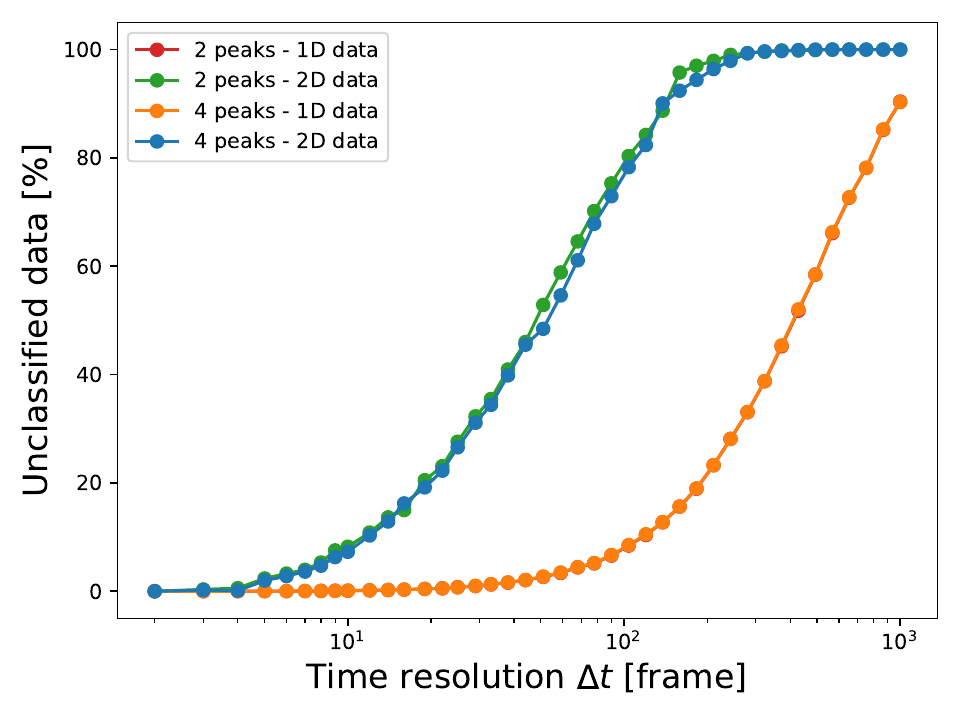}
    \caption{The fraction of unclassified data points in the four Langevin dynamics datasets. As can be seen, by clustering bi-dimensional datasets (blue and green curves) this fraction grows faster with $\Delta t$. Color coding is consistent with Fig.~4 in the main text. Notice that the red and orange curves are tightly superimposed. }
    \label{fig:figs3}
\end{figure}

\newpage

\subsection*{Simulation details}
\subsubsection*{Ice/liquid water coexistence}
An initial configuration for the system is obtained combining configurations of liquid water and $I_h$ ice, each one containing 1024 TIP4P/ICE molecules~\cite{abascal2005potential}. This system was equilibrated for 10~ns at ambient pressure and temperature $T=268$~K. Finally, the production run was performed in $NPT$ conditions, at the same temperature and pressure, for 50~ns, with a sampling time interval of $\delta t = 0.04$~ns. The simulation follows the same protocol illustrated in detail in Ref.~\cite{caruso2023timesoap}. 

\subsubsection*{SOAP power spectra calculation}
At every trajectory frame $t$ along the TIP4P/ICE MD trajectories, we computed the SOAP power spectrum for each molecule using the parameters $n_\text{max} = n'_\text{max} = l_\text{max} = 8$ and a cutoff radius $r_c = 10$~\AA. Each SOAP spectrum consists of 576 components. To reduce dimensionality, we performed Principal Component Analysis (PCA), keeping only the first PC for the following analyses. Signals were then denoised using the protocol reported in Ref.~\cite{donkor2024beyond}. 

For the analysis of FIG. 3 in the main text, we also applied the time-lagged Independent Component Analysis (tICA) dimensionality reduction to the same SOAP spectra, using a time-lag of 10 frames (0.4~ns). Results are stable within a broad range of time-lag values. 

\subsubsection*{Bi-dimensional Langevin dynamics}
The Langevin dynamics (in the limit of zero friction) 
\begin{equation}
    x(t + \delta t) = x(t) + \sqrt{2D\delta t}\cdot \eta(t)
\end{equation}
of 100 particles moving in a bi-dimensional plane was simulated using in-house Python code, for 1e5 time-steps, with a time increment $\delta t = 0.01$ and a diffusion coefficient $D = 0.6$, with $\eta(t)$ random Gaussian noise. 
The free energy landscapes have two and four minima, located in (0, 0), (0, 1) and (0, 0), (0, 1), (1, 0), (1, 1) respectively, and are defined as 
\begin{equation}
    U(x, y) = -\ln{\left[\exp\left(-\frac{x^2 + y^2}{2\sigma^2}\right) + \exp\left(-\frac{x^2 + (y - 1)^2}{2\sigma^2}\right)\right]}
\end{equation}
(and analogously for the 4 minima version, and for the three-dimensional version) with $\sigma = 0.12$. 

\subsection*{Methods details}
\subsubsection*{Static dataset clustering}
The results shown in Fig.~1d in the main text are obtained using onion clustering with $\Delta t = \tau = 50$~ns, thus clustering time-series as belonging on one of the two main environments (ice, in blue, and liquid, in orange) or unclassified (in gray). Similar results can be obtained using any Gaussian Mixture Model with $k = 2$ components on the flatten dataset. 

\subsubsection*{Shannon entropy calculations}
Given a dataset of measured values $\{x_i\}$, with $i\in\{0, ..., N - 1\}$, we construct a histogram $\{b_j, p_j\}$, where $b_j$ are the bins and $p_j$  are the normalized counts ($\sum_j p_j = 1$). Notice that to build the histogram, the number of bins $n_b$ has to be chosen. The Shannon entropy of the dataset is then 

\begin{equation}
    H = -\sum_{j=0}^{n_b} p_j\log_2 p_j
    \label{eq:h_data}
\end{equation}

After the data are clustered with a suitable algorithm, obtaining $K$ clusters, each containing values $\{x_i\}_k$, the same procedure is applied to compute each cluster’s entropy $H_k$, using the same binning $\{b_j\}$ as for the full dataset. Using Eqs.~(2) and~(3) in the main text, we can then compute the information gain as 

\begin{equation}
    \Delta H = H - \sum_{k=1}^{K}f_k H_k
    \label{eq:info_data}
\end{equation}

Notice that in a trivial clustering where all data points belong to a single cluster, $\Delta H = 0$, as expected. Similarly, for a random clustering that distributes points into $K$ clusters with identical distributions, the information gain vanishes on average. 
Throughout the text, all Shannon entropy are normalized between 0 and 1 by diving them by the maximum possible entropy, $\log_2 n_b$. A number of bin $n_b = 20$ was used; different number of bins, up to $n_b = 40$, were tested, giving qualitatively comparable results. 
All the code used to perform the analyses for this work is based on the Dynsight software~\cite{dynsight}. 

\subsubsection*{Onion clustering}
Onion Clustering~\cite{becchi2024layer} is a recently introduced method for single-point time series clustering. It is specifically designed to detect stable patterns within noisy temporal data. The method depends on a single parameter, the time resolution $\Delta t$, which defines the shortest duration a segment must have to be considered part of a stable state. This parameter enforces temporal coherence by ensuring that only sequences persisting for at least $\Delta t$ frames are assigned to a specific state. Segments that do not meet this criterion are grouped into an ``unclassified'' cluster. 

The clustering operates through an iterative mechanism: 

\begin{enumerate}
    \item A Gaussian distribution is fitted to the most prominent peak in the current probability density function of the data. 
    \item All sequences whose values fall within a predefined proximity to the Gaussian mean are attributed to the corresponding state. 
    \item These sequences are then excluded from the dataset, modifying the remaining density landscape. 
    \item The process repeats with the updated dataset. 
\end{enumerate}
Iterations continue as long as unclassified sequences remain and can still be associated with a Gaussian distribution. Eventually, each segment is either assigned to a Gaussian-defined state or left in the unclassified group.

\end{document}